# Atomic moments in $Mn_2CoAl$ thin films analyzed by X-ray magnetic circular dichroism

M.E. Jamer[1], B.A. Assaf[1], G.E. Sterbinsky[2], D.A. Arena[2] and D. Heiman[1]

[1]Department of Physics, Northeastern University, Boston, MA 02115 USA
[2]Photon Sciences Directorate, Brookhaven National Laboratory, Upton, NY, 11973 USA

**Abstract**

Spin gapless semiconductors are known to be strongly affected by structural disorder when grown epitaxially as thin films. The magnetic properties of $Mn_2CoAl$ thin films grown on GaAs (001) substrates are investigated here as a function of annealing. This study investigates the atomic-specific magnetic moments of Mn and Co atoms measured through X-ray magnetic circular dichroism as a function of annealing and the consequent structural ordering. The results indicate that the structural distortion mainly affects the Mn atoms as seen by the reduction of the magnetic moment from its predicted value.

**Introduction**

Spin gapless semiconductors (SGS) are novel materials that combine both half-metallic and semiconducting properties.[1-3] SGS materials are being investigated for spintronic devices due to their unique magnetic and electrical properties. In Fig. 1(a), a schematic of the density of states is shown, where the minority spin state acts as a half metal, whereas the majority spin state acts like a zero-gap semiconductor. These compounds are predicted to have high spin-polarization of carriers and are capable of switching between spin-polarized electrons and holes through tuning the Fermi energy. These compounds also show a high Curie temperature (600-800K), which makes them well-positioned for room temperature devices. Several predicted SGS compounds have been made in bulk form, such as $Mn_2CoAl$ and $Cr_2CoGa$.[4,5] $Mn_2CoAl$ thin films have been shown to grow epitaxially on GaAs (001) substrates,[6,7] though they have shown disorder that affects the SGS and magnetic properties.[8,9]

X-ray diffraction (XRD) measurements on thin films of $Mn_2CoAl$ showed that the samples grew epitaxially with a tetragonal distortion.[6] Upon annealing, the epitaxial and tetragonally-distorted lattice transforms into an oriented cubic cell. In this study, the magnetic properties of $Mn_2CoAl$ thin films were investigated at various stages of annealing via magnetic synchrotron techniques. The X-ray magnetic circular dichroism (XMCD) spectra were taken at room temperature with a 1.5 T applied field using electron absorption detection. The beam was circularly polarized ~70 %. Analysis of the XMCD spectra led to measurements of the atom-specific moments of the Mn and Co atoms.

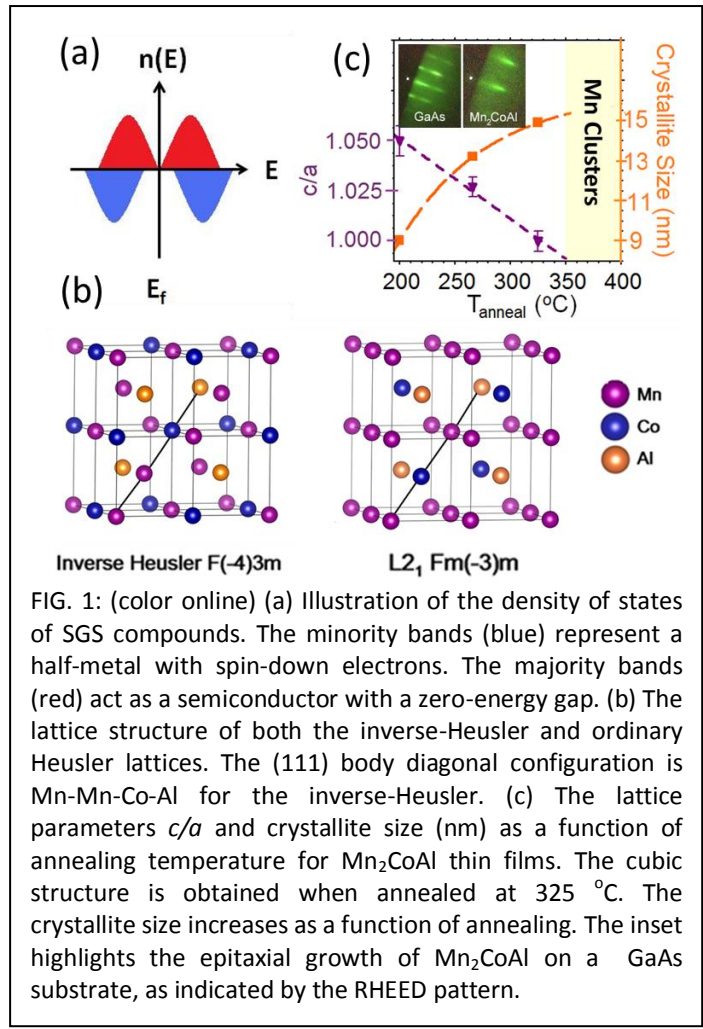

FIG. 1: (color online) (a) Illustration of the density of states of SGS compounds. The minority bands (blue) represent a half-metal with spin-down electrons. The majority bands (red) act as a semiconductor with a zero-energy gap. (b) The lattice structure of both the inverse-Heusler and ordinary Heusler lattices. The (111) body diagonal configuration is Mn-Mn-Co-Al for the inverse-Heusler. (c) The lattice parameters $c/a$ and crystallite size (nm) as a function of annealing temperature for $Mn_2CoAl$ thin films. The cubic structure is obtained when annealed at 325 °C. The crystallite size increases as a function of annealing. The inset highlights the epitaxial growth of $Mn_2CoAl$ on a GaAs substrate, as indicated by the RHEED pattern.

**Experimental Techniques**

The $Mn_2CoAl$ thin films (~70 nm thick) were grown using solid source molecular beam epitaxy (MBE) on GaAs (001) oriented substrates. The GaAs substrates were desorbed at 620 °C in an As flux ($10^{-5}$ Torr) to remove the surface oxide layer. Reflection high energy electron diffraction (RHEED) was used to observe the surface crystallinity after desorption and during the growth process. The RHEED patterns show a well-ordered surface crystallinity of the substrate after desorption and a coherent pattern of the film after growth, indicating that $Mn_2CoAl$ grew epitaxially with respect to the substrate. The films were capped with 2 nm of Al to form a 3 nm $AlO_x$ barrier to prevent oxidation. The samples were subsequently annealed at 200-400 °C for ~30 minutes in a $10^{-8}$ Torr vacuum.

The averaged magnetic properties were measured using superconducting quantum interface device (SQUID) magnetometry. Atomic-specific magnetic moments of the Mn and Co atoms were obtained by analyzing the XMCD polarization-dependent X-ray absorption spectra (XAS) of the Mn and Co transitions, taken using the U4B beamline at the National Synchrotron Light Source at the Brookhaven National Laboratory taken in total electron yield (TEY) mode. The field-dependent measurements were taken at various energies, including energies above, below and at the $L_2$ and $L_3$ edges of the Mn and Co atoms. The penetration depth of the X-rays estimated from the peak absorption is 40-50 nm, indicating that the spectra represent the bulk of the film.

**XRD Results**

The $X_2YZ$ inverse-Heusler structure (XA) with space group F(-4)3m is associated with the SGS compounds. This lattice has 4 *fcc* sublattices, with a 4-atom basis configuration X-X-Y-Z along the body diagonal, shown in Fig. 1(b).[10] The inverse-Heusler structure (left) is seen to be similar to the ordinary Heusler $L2_1$ structure (right). The ordinary Heusler compounds such as $Co_2MnAl$ ($X_2YZ$) are identified with space group Fm(-3)m,[11] but with a 4-atom basis configuration of X-Y-X-Z along the body diagonal.

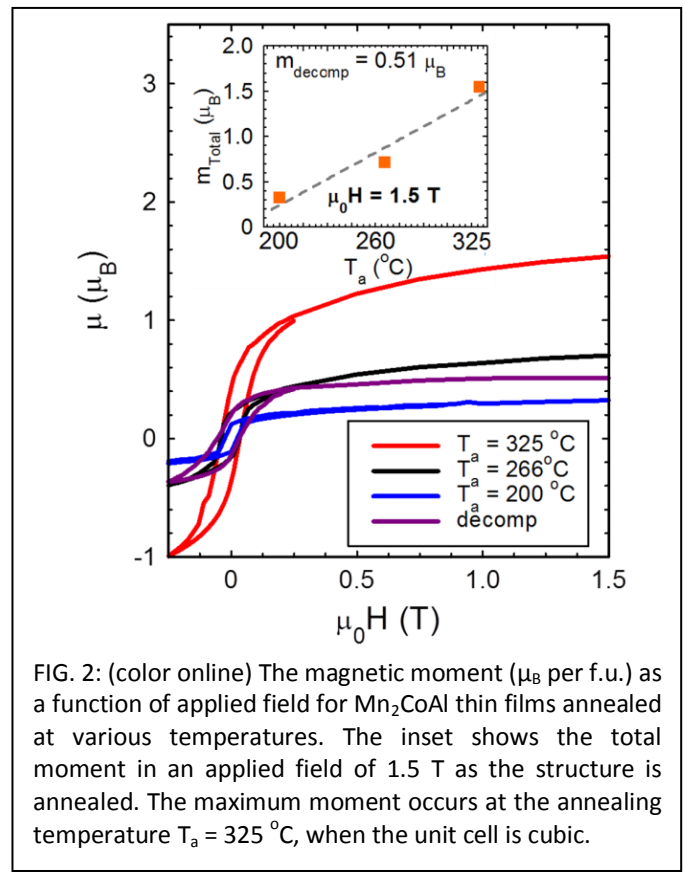

FIG. 2: (color online) The magnetic moment ($\mu_B$ per f.u.) as a function of applied field for $Mn_2CoAl$ thin films annealed at various temperatures. The inset shows the total moment in an applied field of 1.5 T as the structure is annealed. The maximum moment occurs at the annealing temperature $T_a$ = 325 °C, when the unit cell is cubic.

XRD was used to characterize the annealing effects on the structural properties of $Mn_2CoAl$. The XRD Bragg patterns were obtained using a Cu-Kα radiation source with wavelength 1.5814 Å. Since the films were aligned along the (00*L*) plane, in-plane diffraction was used to determine the *a* lattice parameter. Figure 1(c) shows the *c/a* ratio as a function of the annealing temperature. The data shows that the as-grown structure has the largest tetragonal distortion (*c/a* = 1.05) and the structure annealed at $T_a$ = 325 °C for 30 minutes has a cubic structure (*c/a* = 1.00). When the sample is annealed at higher temperatures, 400 °C, the Mn atoms coalesce in the lattice to form Mn crystallites that are observed as a Mn peak in the XRD pattern.6 When the lattice first becomes cubic at $T_a$ = 325 °C, the chemical ordering parameter is *S* = 0.45, calculated from the measured and theoretical intensities, indicating that there is atomic swapping. The crystallite size is computed from the Scherrer formula using the width of the (004) peaks, shown in Fig. 2(inset). It is seen that the crystallite size increases linearly with increasing annealing temperature. The crystallite size in the *c*-axis direction is affected by the *c*-axis straining to match the GaAs substrate, and relaxing as a function of thickness. Previous theoretical calculations determined that the tetragonality of the lattice has almost no effect on the band structure properties, and little effect on the atomic moments; however, in the case of atomic swapping, the magnetic and band structure properties are highly affected.[9] It is noted that both tetragonal distortion and atomic swapping in the thin films impact the exchange coupling of the magnetic moments. The magnetic results will be quantified by $T_a$, the annealing temperature.

**SQUID Magnetometry**

SQUID magnetometry was used to measure the room temperature total magnetic moment of $Mn_2CoAl$ thin films that were annealed between 200 and 400 °C. Figure 2 shows the field dependence of the magnetic moment up to 1.5 T, in units of Bohr magneton ($\mu_B$) per formula unit (f.u.). It is seen that the moment does not saturate at these fields. The inset in Fig. 2 shows that the magnetic moment measured at 1.5 T increases as the structure is annealed from 200-325 °C. The moment reaches a value of 1.54 $\mu_B$/f.u. when the sample is cubic ($T_a$ = 325 °C). The moment decreases when the structure is further annealed (350-400 °C), coinciding with the segregation of Mn into crystallites in the lattice.[6] It has been noted that due to atomic disorder, the magnetic moment does not saturate up to fields of 5 T, which indicates the overall moment has not reached the predicted saturation of 2.0 $\mu_B$/f.u.[2,3] The atomic disorder is expected to lead to incoherent coupling of the Mn atoms, resulting in a reduced moment that affects the Co moment.

**Measurements of Atomic Moments**

The spin and orbital magnetic moments of Mn and Co were measured through analysis of the XAS taken at the $L_2$ and $L_3$ edges with circularly polarized X-rays.[12,13] Figure 3(a) shows the XAS of the $L_3$ and $L_2$ transitions for the Mn and Co atoms, for $T_a$ = 200-325 °C and for the partially decomposed film.[14] XMCD probes the element-specific magnetic vector moment <m> for a specific orbital energy transition.[15] The XMCD patterns reveal the dichroic differences between the left and right circularly polarized XAS patterns. Figure 3(b) shows an overall trend in the XMCD patterns. As $T_a$ increases, the Mn signal increases monotonically, while the Co signal develops a minimum at $T_a$ = 260 °C.

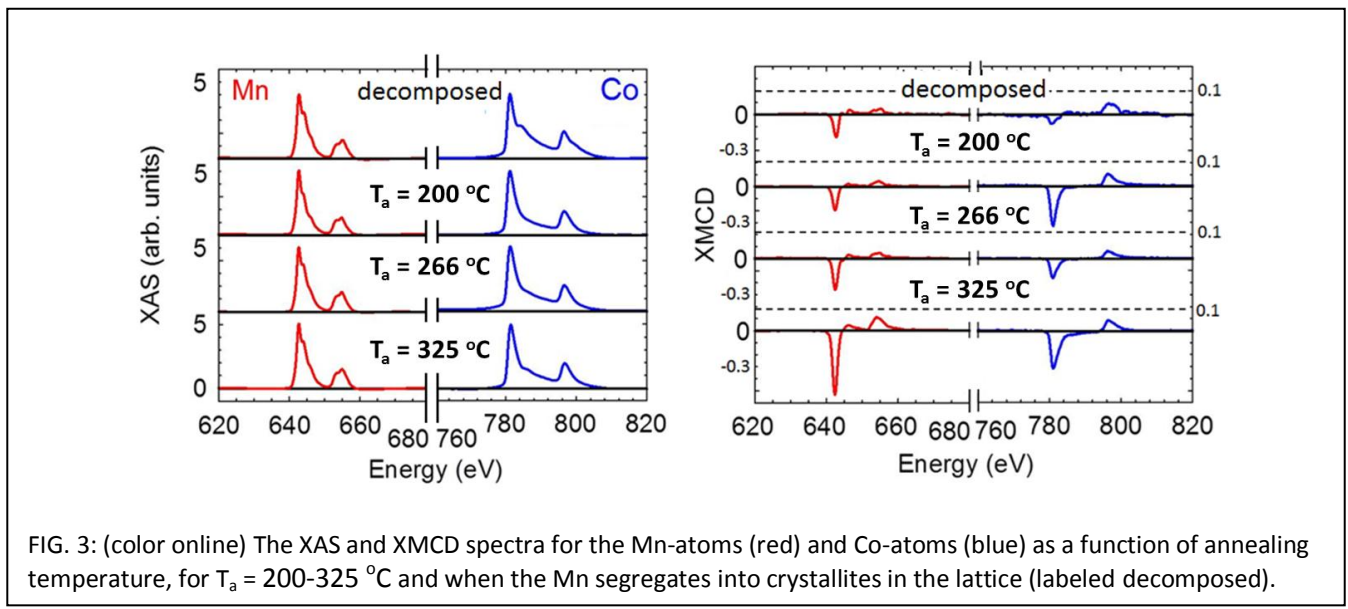

FIG. 3: (color online) The XAS and XMCD spectra for the Mn-atoms (red) and Co-atoms (blue) as a function of annealing temperature, for $T_a$ = 200-325 °C and when the Mn segregates into crystallites in the lattice (labeled decomposed).

The trends in magnetic moment as a function of $T_a$ are more apparent in Fig. 4, showing the separation of spin and orbital moments of Mn (red) and Co (blue) that are derived from the XMCD patterns. The spin moments are larger than the orbital moments as expected. Note here that the errors in the magnetic moments can deviate by as much as 10-30 % depending on the number of atoms in the $d$-valence shell.[16] The first thing to note is the decrease in Co moments at $T_a$ = 266 °C, however, the orbital moment of the Mn atom increases at this point as well. The large decrease in the Co spin moment at $T_a$ = 266 °C, with an increase in the Mn orbital moment indicates that there could be some switching in the Mn and Co atomic sites.[17] Second, there is a general trend where the atomic moments are largest in the cubic phase ($c/a$ = 1.00), and nearly vanish when the film begins to decompose and Mn clusters develop. Figure 4(b) shows the Co magnetization from hysteresis scans scaled from TEY to the magnetic moment found through XMCD analysis. The Co moment does not saturate at ~1.8 T for $T_a$ = 200-325 °C. Table 1 summarizes these measurements and shows the comparison between the SQUID and XMCD results. The value $\mu_{Tot}$ corresponds to the addition of the orbital and the spin moments from the XMCD results.[18] The SQUID and XMCD values are reasonably close except for the values at $T_a$ = 200 °C. The other total moment values are comparable to the XMCD results within error, as there is a large error associated with XMCD measurements[16].

The valence states of the Mn and Co d-orbitals were found through modeling the XAS.[19] Analyzing the Co XAS patterns led to identifying the valence state as $Co^{3+}$ ($3d^6$) when $T_a$ = 200-325 °C. However, when the lattice decomposes at the highest annealing temperature, 400 °C, a secondary valence state emerges, leading to a combination of $Co^{3+}$ and $Co^{4+}$. Similarly, it was found that analysis of the Mn XAS curves lead to a constant valence state for $T_a$ = 200-325 °C, corresponding to $Mn^{2+}$ ($3d^5$).

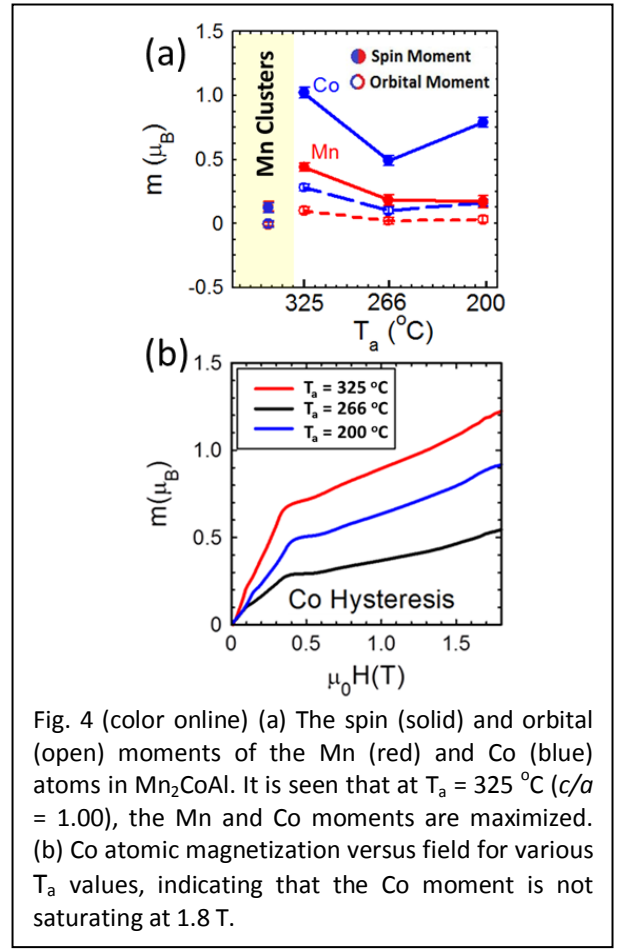

Fig. 4 (color online) (a) The spin (solid) and orbital (open) moments of the Mn (red) and Co (blue) atoms in $Mn_2CoAl$. It is seen that at $T_a$ = 325 °C ($c/a$ = 1.00), the Mn and Co moments are maximized. (b) Co atomic magnetization versus field for various $T_a$ values, indicating that the Co moment is not saturating at 1.8 T.

TABLE 1: The spin magnetic moments of $Mn_2CoAl$ measured by SQUID magnetometry ($m_{Tot}$) and the atom-resolved moments in $\mu_B$ at 1.5 T versus the annealing temperature $T_a$. The crystallite size (D nm) and $c/a$ ratio are also tabulated. The next-to-last column lists the sum of the Mn and Co moments ($\mu_{Tot}$), and the last column lists the SQUID results for comparison.

| $T_{anneal}$ | D | $c/a$ | $m_{s(Mn)}$ | $m_{s(Co)}$ | $m_{orb(Mn)}$ | $m_{orb(Co)}$ | $\mu_{Tot}$ | $m_{Tot}$ |
|---|---|---|---|---|---|---|---|---|
| 200 | 9.1 | 1.05 | 0.15 | 0.8 | 0.02 | 0.16 | 1.13 | 0.33 |
| 266 | 13.2 | 1.03 | 0.16 | 0.47 | 0.08 | 0.09 | 0.80 | 0.71 |
| 325 | 14.9 | 1.00 | 0.38 | 0.9 | 0.06 | 0.28 | 1.62 | 1.54 |
| 400 | - | - | 0.13 | 0.21 | 0 | 0 | 0.34 | 0.51 |

**Conclusions**

$Mn_2CoAl$ thin films grown on GaAs substrates have varying amounts of disorder depending on the annealing temperature, which leads to changes in the magnetic properties. SQUID magnetometry shows that the magnetic moment increases as the annealing temperature increases which corresponds to the tetragonal to cubic transition. At higher annealing temperatures above 350 °C, Mn crystallites form in the crystal structure. The magnetic moments of the Mn and Co atoms were measured using XMCD. The moment of the Co atoms in the proper cubic structure is as-predicted; however, the moment of the Mn atoms is smaller than the predicted value due to disorder. These results indicate a crucial need for further theoretical studies of the magnetic properties and band structure when these spin gapless semiconductors are compositionally and structurally disordered.[9,20,21]

**Acknowledgements**

We thank T. Devakul for his work on the samples. M.E.J. acknowledges M. Loving's advice on XMCD analysis. The work was supported by the National Science Foundation grants DMR-0907007 and ECCS-1402738. Use of the National Synchrotron Light Source, Brookhaven National Laboratory, was supported by the U.S. Department of Energy, Office of Science, Office of Basic Energy Sciences, under Contract No. DE-AC02-98CH10886.